\documentclass[lettersize,journal]{IEEEtran}
\usepackage{amsmath,amsfonts}
\usepackage{algorithmic}
\usepackage{algorithm}
\usepackage{array}
\usepackage[caption=false,font=normalsize,labelfont=sf,textfont=sf]{subfig}
\usepackage{textcomp}
\usepackage{stfloats}
\usepackage[hidelinks]{hyperref}
\usepackage{verbatim}
\usepackage{graphicx}
\usepackage{epstopdf}
\usepackage{cite}
\usepackage{url}
\usepackage[hidelinks]{hyperref}
\begin{document}

\title{Generative Channel Knowledge Base With Environmental Information for Joint Source-Channel Coding in Semantic Communications}

\author{Xudong Long, \IEEEmembership{Student Member, IEEE,} Hao Chen, \IEEEmembership{Member, IEEE,} Dan Wang, \IEEEmembership{Member, IEEE,} Chen Qiu, \IEEEmembership{Member, IEEE,}  Nan Ma, \IEEEmembership{Member, IEEE,} Xiaodong Xu, \IEEEmembership{Senior Member, IEEE,} Yubin Zhao, \IEEEmembership{Senior Member, IEEE,}

\thanks{Xudong Long and Yubin Zhao are with the School of Microelectronics Science and Technology, Sun Yat-Sen University, Zhuhai, China, 519082, and Xudong Long also with the Department of Broadband Communication, Pengcheng Laboratory, Shenzhen, China, 518055. Email: longxd@mail2.sysu.edu.cn, zhaoyb23@mail.sysu.edu.cn.}

\thanks{H. Chen, D. Wang, C. Qiu, Nan Ma and Xiaodong Xu are with the Department of Broadband Communication, Pengcheng Laboratory, Shenzhen, China, 518055, Email: chenh03@pcl.ac.cn, wangd01@pcl.ac.cn, qiuch@pcl.ac.cn.}
\thanks{ (Corresponding authors: Hao Chen and Yubin Zhao)}


}

\markboth{Journal of \LaTeX\ Class Files,~Vol.~14, No.~8, August~2021}%
{Shell \MakeLowercase{\textit{et al.}}: A Sample Article Using IEEEtran.cls for IEEE Journals}


\maketitle

\begin{abstract}
Semantic knowledge bases are regarded as a promising technology for upcoming 6G communications. However, existing studies mainly focus on source-side semantic modeling while overlooking the structural impact of propagation environments on semantic transmission performance. To address this issue, we propose a generative channel knowledge base (CKB) with environmental information to facilitate joint source-channel coding (JSCC) in semantic communications (SemCom) systems. First, to enable the construction of the CKB, an environment-aware dataset is established by collecting spatial position information, global image features, fine-grained semantic features, and the corresponding channel matrices. A region-of-interest (ROI)-based filtering algorithm is further designed to remove semantic components that are irrelevant to signal propagation. Second, a Transformer-based generative framework is developed to learn the mapping between multidimensional environmental information and channel matrices. A self-attention mechanism is introduced to adaptively fuse heterogeneous features, enabling the construction of a structured CKB. Third, a CKB-driven JSCC SemCom architecture is proposed, where the generated channel knowledge is injected into both of the encoder and decoder to jointly exploit source semantics and channel-environment priors in an end-to-end manner. Experimental results demonstrate that the proposed multidimensional feature fusion method achieves a channel matrix estimation error at the $10^{-3}$ level. Moreover, the CKB-driven JSCC SemCom framework integrated into SemCom systems significantly outperforms existing benchmark schemes in terms of transmission performance.
\end{abstract}

\section{Introduction}
With the advancement of artificial intelligence (AI) and sixth-generation (6G) wireless communication technologies, semantic communications (SemCom) has been recognized as a potential paradigm for next-generation mobile networks, which aims to transmit semantic representations that directly support downstream tasks e.g., classification and reconstruction  \cite{shi2021semantic,gunduz2022beyond}. By enabling the encoding and transmission of high-level semantic information, SemCom facilitates a wide range of emerging intelligent applications, including smart manufacturing, autonomous driving, and intelligent transportation systems \cite{huo2024image}. However, when SemCom is deployed in practical 6G environment, e.g., vehicle-to-everything (6G-V2X) scenarios, its transmission performance can be significantly affected by the complex wireless propagation environment. Due to the high mobility of vehicles and users, blockage effects in complex traffic environments, multipath propagation, and the dense presence of traffic participants, communication links are characterized by pronounced time variability, strong nonlinearity, and high uncertainty \cite{yang2022semantic}. In such scenarios, wireless channel characteristics are influenced by environmental factors, including building geometry, spatial distribution, and multipath interactions, which lead to aggravated signal fading and degraded channel state information (CSI), thereby undermining the reliability of end-to-end SemCom transmission.

To cope with these challenges, the concept of a semantic knowledge base (SKB) has been introduced into SemCom systems, where prior knowledge is utilized for inference and recovery, thereby improving the efficiency of SemCom \cite{gao2025agentic}. To the best of our knowledge, existing research has predominantly focused on source-side semantic modeling and knowledge representation, whereas the impact of wireless propagation environments on SemCom performance has not yet been systematically investigated \cite{li2025semantic}. In practical wireless communication scenarios, signal propagation is jointly governed by multiple environmental factors, including spatial structures, obstacle distributions, and scene-level semantics. These factors not only affect wireless channel characteristics but may also be intrinsically correlated with the semantic information embedded in the source data \cite{gimenez2024semantic}. Consequently, propagation characteristics in real environments cannot be adequately captured by conventional statistical channel models or simplified channel assumptions, which further limits the performance of SemCom systems in complex scenarios.

Meanwhile, to address the impact of environmental factors on communication system performance in 6G-V2X scenarios, existing studies have primarily focused on three categories of approaches, namely wireless environment information modeling \cite{zhang2023channel}, channel knowledge maps (CKM) \cite{li2023channel}, and radio maps \cite{zhou2025generative}.
By integrating multi-source environmental information, such as geographical location, environmental structures, and historical measurement data, wireless channel characteristics can be modeled and predicted \cite{shen2025efficient}. Therefore, the channel time variability and uncertainty induced by high mobility, blockage effects, and multipath propagation in 6G-V2X scenarios are alleviated to a certain extent, whereby the accuracy of channel state information acquisition is improved and the reliability of communication links is enhanced \cite{ren2025channel}.

However, the aforementioned studies are primarily oriented toward conventional communication system design and fail to provide an environmental knowledge representation that can directly support semantic representation, semantic encoding, and semantic reasoning. Consequently, they are insufficient to meet the requirements of  SemCom systems in terms of environmental awareness and semantic collaboration. Against this backdrop, incorporating environmental information into semantic transmission remains challenging in several aspects. 
First, existing environment modeling approaches are predominantly based on either "store-match-interpolate" or physics-driven paradigms, which hinder the characterization of mapping relationships between multi-dimensional environmental information and high-dimensional channel states within a unified learning framework \cite{feng2025recent}. 
Second, current methods mainly focus on knowledge storage, retrieval, and representation while lacking generative inference capabilities for unseen scenarios \cite{liu2025channel}. 
Third, most existing JSCC-based SemCom schemes are developed under simplified or known channel models, and a unified end-to-end framework that jointly integrates environmental information, channel knowledge, and semantic transmission remains absent \cite{huang2025adaptive}.

To address the aforementioned issues, a multidimensional environment-aware CKB is proposed for 6G-V2X SemCom systems, along with a CKB-driven JSCC SemCom architecture. Specifically, an environment-aware dataset integrating multidimensional environmental information and channel features is constructed to support CKB modeling. By incorporating environment-aware channel knowledge into the SemCom process, the proposed framework enables more accurate channel representation and improved transmission reliability in complex propagation environments.
The main contributions of this paper are summarized as follows:
\begin{itemize}
	\item First, to enable the development of the CKB, a multidimensional environmental dataset for 6G-V2X scenarios is constructed and released, which includes user spatial position information, coarse-grained global image features, fine-grained semantic features, and the corresponding channel matrices. Furthermore, a region-of-interest (ROI)-based filtering algorithm is designed to remove semantic components within the communication user (CU) region that are irrelevant to signal propagation.
	

	\item Second, a generative CKB is constructed to model the mapping between multidimensional environmental information and channel matrices based on a Transformer network, thereby enabling the quantification of environmental impacts on SemCom system performance. By introducing a self-attention mechanism, adaptive fusion of heterogeneous multidimensional environmental features is achieved, allowing the model to dynamically capture the contributions of different environmental factors to channel characteristics, thereby improving the effectiveness of environmental information modeling.	
	
	\item Third, a CKB-driven JSCC SemCom architecture is proposed, in which the channel knowledge corresponding to multidimensional environmental information is retrieved from the CKB to characterize rich environmental spatial features and is integrated into the JSCC process for end-to-end training. Compared with conventional DeepJSCC schemes, the proposed method can more accurately characterize channel conditions, thereby significantly improving transmission performance and system robustness in complex 6G-V2X environments.
\end{itemize}

Simulations are conducted in a realistic 6G-V2X environment using integrated simulation platforms with multidimensional environmental information and channel data. The proposed method is compared with conventional DeepJSCC and JSCC without CKB. Results show that the mean square error (MSE) of the proposed framework reaches $10^{-3}$ and our system significantly improves transmission performance in terms of peak signal-to-noise ratio (PSNR) and structural similarity index (SSIM) across a wide signal-to-noise ratio (SNR) range. Furthermore, the results demonstrate that incorporating environment-aware channel knowledge effectively mitigates channel estimation errors and enhances the robustness and generalization capability of SemCom systems in complex propagation environments. 

The rest of this paper is organized as follows: Section II reviews related work. Section III presents the system model. Section IV introduces the environmental CKB. Section V presents the CKB-driven JSCC SemCom framework. Section VI provides the simulations. Section VII concludes the whole paper.

\section{Related Work}
Benefiting from the advancement of AI technologies, conventional communication systems have progressively converged with learning-based methodologies, thereby facilitating the emergence and evolution of SemCom paradigms. By projecting source information (SI) into a compact low-dimensional feature space and directly conveying the most informative feature representations, multiple-input multiple-output (MIMO) SemCom systems enable accurate information reconstruction at the receiver via a semantic decoder, thus supporting efficient information delivery in 6G-V2X scenarios \cite{dai2022nonlinear}. He et al. introduced an adaptive encoding mechanism that allocates different channel rates to different data modalities based on feature importance \cite{he2023rate}. Huang et al. proposed a generative SemCom framework, wherein efficient data reconstruction is proposed by transmitting only the SI semantic key features associated with semantic labels through a semantic filtering approach \cite{huang2025visual}. Shao et al. proposed a SemCom scheme under power constraints, which attains excellent image reconstruction results through three peak power ratio restoration techniques \cite{shao2022semantic}. Bou et al. proposed a JSCC scheme in which the SI and complex-valued channel information are directly embedded into the encoding process, thereby gaining high-performance data transmission and image restoration with low  SNR \cite{bourtsoulatze2019deep}. Although deep-learning-based SemCom systems facilitate efficient extraction and transmission of semantic information, semantic reference and interpretation become fundamentally difficult to preserve in a consistent and aligned manner when the prior knowledge structures and semantic reference frameworks at the transmitter and the receiver are misaligned, leading to semantic misinterpretation and degraded reliability of semantic reconstruction, and ultimately constraining the overall system performance.

To overcome the challenges of accurately modeling environment-dependent wireless channels and the resulting performance degradation of SemCom systems in complex propagation environments, prior knowledge has been introduced into SemCom systems through the construction of a SKB, which facilitates semantic encoding, semantic inference, and semantic reconstruction. Ren et al. explored the fundamental concept of a knowledge base, where key information is extracted from SI for transmission and efficiently reconstructed at the receiver by exploiting the shared knowledge between the transmitter and the receiver \cite{ren2024knowledge}. Sun et al. and Zhou et al. proposed effective methodologies for constructing knowledge bases aimed at improving the efficiency of semantic transmission \cite{sun2023semantic,zhou2024moc}.
Wang et al. employed a compact SKB composed of semantic knowledge vectors corresponding to image categories, together with a multi-layer feature extractor, to efficiently perform semantic tasks \cite{wang2023knowledge}.
Peng et al. proposed a shared knowledge base to support source transmission, where the system integrates relevant prior knowledge to infer the remaining information with the assistance of the knowledge base. Then, less symbols are required for transmission without sacrificing semantic expressiveness \cite{yi2023deep}.
Li et al. developed a cooperative knowledge base framework, in which the knowledge bases are iteratively updated through inter-user interactions, enabling significant performance gains with extremely low transmission overhead \cite{li2024cooperative}.
Hu et al. introduced an electronic codebook based on a knowledge base, where prior side information is provided to enhance the effectiveness of semantic encoding \cite{hu2023robust}.
However, current knowledge-base-related research mainly concentrates on source knowledge, while the impact of dynamic environmental factors in 6G-V2X scenarios, including high mobility, blockage, and multipath propagation, remains insufficiently explored, thereby presenting new challenges for further enhancing the performance of SemCom systems.

In conventional communication systems, by analyzing environmental information in 6G-V2X scenarios, e.g., geographical layouts, building structures, and historical measurements of traffic participants, the characteristics of wireless channels can be modeled and predicted, thereby alleviating the temporal variability and uncertainty induced by environmental factors. \cite{xu2023deep,wu2024deep}. Wen et al. determined the optimal beamforming by analyzing image data heatmaps in point-to-point communication, thus facilitating channel link construction \cite{wen2023vision}. Charan et al. proposed a vision-assisted environmental perception scheme that utilizes deep learning to analyze scattering characteristics in images for channel prediction in SemCom systems \cite{charan2023millimeter}. Sun et al. proposed a channel prediction method by establishing the correlation between propagation environments and channel states through graph neural networks \cite{sun2022environment}. Moreover, CKM is proposed to describe channel characteristics across different environments by integrating all channel-related information between transmitters and receivers within the communication environment \cite{zeng2024tutorial}, such as channel gain maps \cite{dall2011channel}, power spectral density maps \cite{teganya2021deep}, line-of-sight (LoS) probability maps \cite{zeng2023ckm}, 3D point map \cite{wang2025point}, communication rate maps \cite{zheng20205g}, and coverage maps \cite{kasparick2015kernel}. These maps enhance the environmental perception capability of communication systems and reduce the need for complex real-time channel state information acquisition \cite{zeng2021toward}. However, these methods rely on scalar metrics and single-dimensional environmental information, which cannot adequately characterize complex propagation environments. Moreover, a dedicated CKB that explicitly models the mapping between environmental information and channel characteristics for SemCom systems has not yet been established. Motivated by this gap, we construct a multidimensional environmental dataset and further establish a CKB by learning the mapping between environmental information and channel matrices. Based on this foundation, a CKB-driven JSCC SemCom architecture for SemCom systems is proposed to enable environment-aware semantic transmission.

\begin{figure}[t]
	\begin{center}
		\includegraphics[width=3.5in]{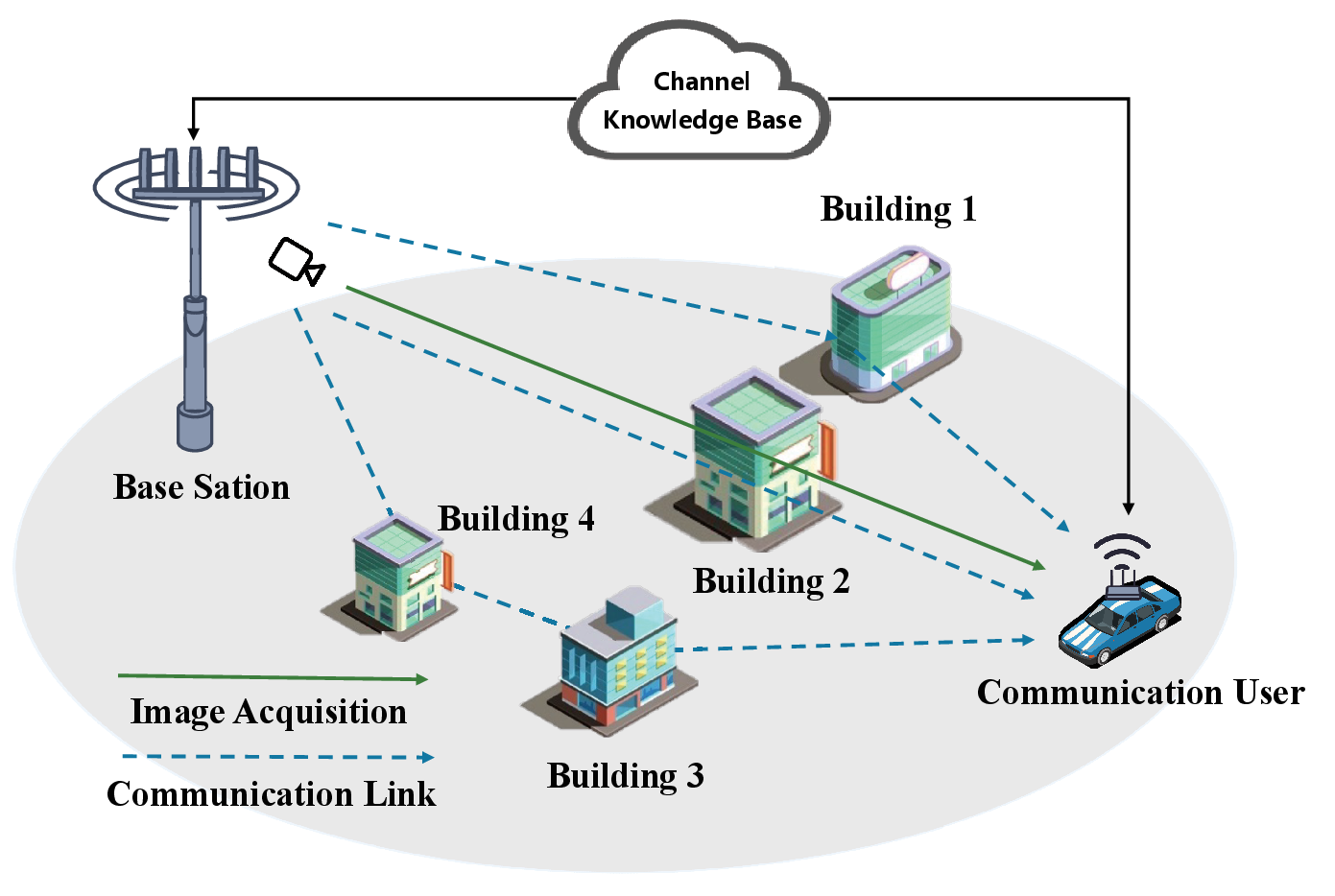}
		\caption{Generative CKB with environmental information for the SemCom system.}
 		\label{fig1_1}
	\end{center}
\end{figure}

\section{System Model}
As depicted in Fig. \ref{fig1_1}, we consider a SemCom system with one  base station (BS), one CU, and multiple buildings. The BS is equipped with an $M$-element antenna array and an environmental sensing module  for signal transmission and image acquisition from the BS-user perspective, respectively. Each CU is equipped with a $K$-antenna array, which is used to receive wireless signals through both line-of-sight (LoS) and non-line-of-sight (NLoS) links. 

During the transmission from the BS to the CU, the signal passes through a real physical environment with multiple buildings, and is influenced by multipath fading, environmental noise, and various interferences. Therefore, a dedicated CKB  is specifically designed for the SemCom system, which reflects the mapping relationship between multidimensional environmental information and channel features. 
Specifically, the SI and the channel knowledge extracted from the CKB, which is obtained according to the mapping between channel characteristics and environmental information, are jointly encoded at the BS transmitter. After propagating through the physical environment influenced by multiple buildings, the CU receiver acquires the corresponding channel knowledge from the shared CKB and performs joint decoding with the received signal to reconstruct the SI.
\begin{figure*}[t]
	\begin{center}
		\includegraphics[width=6.8in]{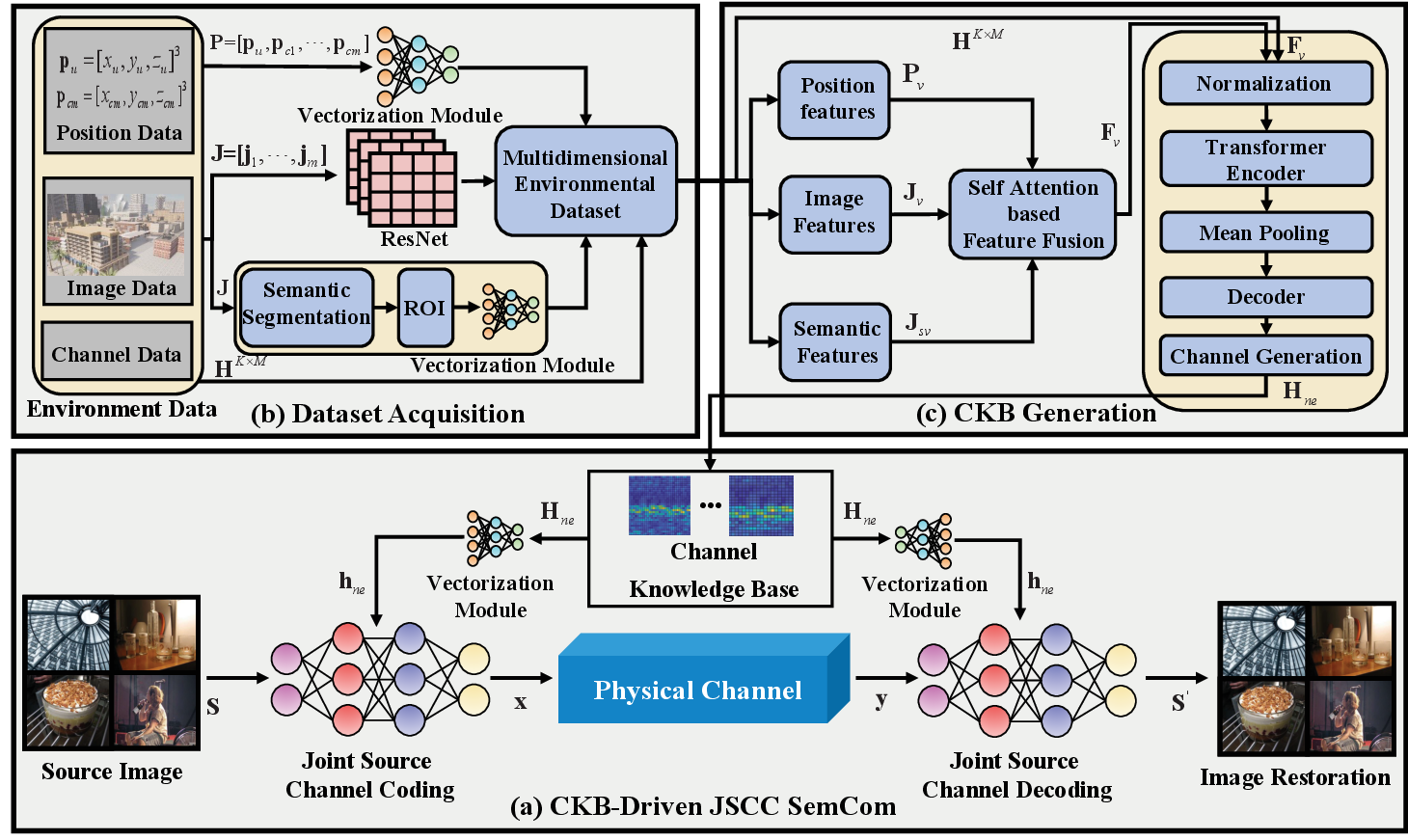}
		\caption{CKB-driven JSCC SemCom framework. (a) CKB-driven JSCC SemCom: channel knowledge corresponding to the environmental information is retrieved from the CKB and integrated into the JSCC framework for end-to-end training. (b) Dataset acquisition: spatial location features, coarse-grained global features, fine-grained local semantic features, and channel data. (c) CKB generation: learning the mapping between multidimensional environmental information and channel matrices using a Transformer-based network. }
		\label{fig1_2}
	\end{center}
\end{figure*}


\subsection{Transmitter}
As shown in Fig. \ref{fig1_2}(a), the BS transmits SI, represented as an image $S \in \mathbb{R}^{h\times w  \times 3} $, over the downlink, where $h$ and $w$ denote the height and width of the image, 3 denotes the RGB channels. Correspondingly,  the CKB is constructed by a finite set of environment-aware channel knowledge entries $\mathbf{\Omega}  = \{\mathbf{H}_{1e},\dots,\mathbf{H}_{ne}\}\in \mathbb{C}^{N \times M \times K}$, where $\mathbf{H}_{ne}$ is generated according to the mapping relationship between the multidimensional environmental information and channel characteristics for point-to-point communication between the BS and the CU, where $N$ denotes the number of communication pairs. In the SemCom system, the CKB is shared between the BS and CU, with the channel knowledge extracted for joint encoding based on the mapping between the information collected by the environmental sensing module and the channel knowledge. Specifically, the channel knowledge $\mathbf{H}_{ne}$ is first transformed into a vector $\mathbf{h}_{ne}$ through a vectorization module, which is then fed into the encoder $\mathcal{S}({\cdot,\cdot})$ for joint encoding with the image data $\mathbf{S}$. Briefly, the transmitter encoder $\mathcal{S}$ maps the channel knowledge vector $\mathbf{h}_{ne}$ onto the image data vector $\mathbf{S}$, producing an output vector $\mathbf{x} \in \mathbb{C}^{M \times 1}$  expressed as:
\begin{equation}
	\begin{aligned}
		\label{outpur_signal}
		\mathbf{x} =\mathcal{S}(\mathbf{S},\mathbf{h}_{ne}).
	\end{aligned}
\end{equation}

\subsection{Channel}
When the signal $ \mathbf{x} $ is transmitted from the BS, the received signal $\mathbf{y} \in \mathbb{C}^{K \times 1}$ at CU is given as:
\begin{equation}
	\begin{aligned}
		\mathbf{y} =\mathbf{H}\mathbf{x}+ \mathbf{n},
	\end{aligned}
\end{equation}
where $\mathbf{H} \in \mathbb{C}^{K \times M}$ denotes the actual physical channel coefficient matrix and $\mathbf{n} \in \mathbb{C}^{K \times 1}$ denotes circularly symmetric complex Gaussian (CSCG) noise, with $\mathbf{n} \sim \mathcal{CN}(0, \sigma\mathbf{I}_{K})$. The channel is modeled via a channel impulse response that captures multipath propagation effects, expressed as: 
\begin{equation}
	\begin{aligned}
		\mathbf{H}(\tau,\boldsymbol{\Theta},\boldsymbol{\Phi})=
		\sum_ma_m\delta(\tau-\tau_m)\delta(\boldsymbol{\Theta}-\boldsymbol{\Theta}_m) \delta(\boldsymbol{\Phi}-\boldsymbol{\Phi}_m),
	\end{aligned}
\end{equation}
where $\tau$, $\boldsymbol{\Theta}$, $\boldsymbol{\Phi}$, and $\sum_ma_m$ denote the time delay, azimuth/elevation angle of departure (AOD), azimuth/elevation angle of arrival (AOA), and the total attenuation coefficient of the $m$-th path, respectively. Each parameter $a_{m}$ is determined by three key factors, the total propagation distance $d_m$, the number of interactions $I_m$, and the attenuation coefficient $\Gamma_m^{(k)}$ per interaction, e.g.,
\begin{equation}
	\begin{aligned}
		a_m=\underbrace{\frac{\lambda}{4\pi d_m}}_{\mathrm{1st}}\cdot\underbrace{\left(\prod_{k=1}^{I_m}\Gamma_m^{(k)}\right)}_{\mathrm{2nd}}\cdot\underbrace{e^{-\mathrm{j}\frac{2\pi}{\lambda}d_m}}_{\mathrm{3rd}},
	\end{aligned}
\end{equation}
where the first term is the free-space attenuation coefficient magnitude, the second term is the attenuation coefficient due to the interaction of the ray with the environment, and the third one represents the phase change caused by free-space propagation. 

\subsection{Receiver}
The received signal $\mathbf{y}$ is processed by the semantic decoder $\mathcal{C}(\cdot,\cdot)$ to obtain the reconstructed image $\mathbf{S}^{'}$. Similar to the encoder, channel knowledge $\mathbf{H}_{ne}$ is extracted from the CKB based on the mapping between environmental information and channel conditions, transformed into a vector $\mathbf{h}_{ne}$ via a vectorization module, and jointly processed with the received signal $\mathbf{y}$ at the decoder:
\begin{equation}
	\begin{aligned}
		\mathbf{S}^{'} =\mathcal{C}(\mathbf{y},\mathbf{h}_{ne}).
	\end{aligned}
\end{equation}

The CKB at the receiver side is identical to that at the transmitter side, ensuring they always share the same background knowledge.

\section{Environmental Channel Knowledge Base}
As illustrated in Fig. \ref{fig1_2}, we propose an environment-aware CKB-driven JSCC framework for SemCom. A multidimensional environmental dataset is first constructed by collecting spatial location information, environmental images, semantic features, and their corresponding channel matrices data. A Transformer-based model is then employed to learn the mapping from environmental representations to wireless channel characteristics, thereby establishing a structured CKB. Based on the learned mapping, environment-relevant channel knowledge is retrieved from the CKB and incorporated into the JSCC encoding and decoding processes, enabling end-to-end optimization through the joint exploitation of source semantics and channel prior knowledge.

\subsection {Multidimensional Environmental Dataset}
We assume that both of the BS and the CU have global positioning system (GPS)-enabled spatial relative localization capability. The resulting localization information is converted into position features through a vectorization module. Meanwhile, an environmental sensing module at the BS captures images from the BS-CU perspective. From these images, global coarse-grained environmental features are extracted, while fine-grained semantic features associated with the CU locality are identified and irrelevant semantic information is filtered out. In addition, the constructed environmental model is imported into an electromagnetic simulation platform, through which channel matrices corresponding to the multidimensional environment are generated. The collection of these one-to-one corresponding data pairs is ultimately organized to form a multidimensional environmental dataset.

$\textbf{Location data:}$ As shown in Fig. \ref{fig1_2}(b), we model the BS as a stationary node positioned at $\mathbf{p}_{u}=[p_{u,x}, p_{u,y}, p_{u,z}] \in \mathbb{R}^{3}$, while the mobile user's locations are denoted by $\mathbf{P}_{c} = [\mathbf{p}_{c1},\cdots,\mathbf{p}_{cN}] \in \mathbb{R}^{N \times 3}$, where $\mathbf{p}_{cN}=[p_{cN,x}, p_{cN,y}, p_{cN,z}]$ defines the localization of the CU in the $N$-th communication pair. The positional domain $\mathbf{P}= [\mathbf{p}_{u},\mathbf{p}_{c1},\cdots,\mathbf{p}_{cN}] \in \mathbb{R}^{(N+1) \times 3}$ represents the set of feasible user locations within the operational environment. With the location as the input, the vectorization module consists of several fully connected (FC) blocks, and each FC block is sequentially stacked by an FC, a BatchNorm, and a ReLU layer. The mathematical function of the vectorization module is denoted as $f_{v}(\mathbf{P},\mathbf{\theta}_{pv})$, i.e.,
\begin{equation}
	\begin{aligned}
		\mathbf{P}_{v} =f_{v}(\mathbf{P},\mathbf{\theta}_{pv}),
	\end{aligned}
\end{equation}
where $\mathbf{P}_{v}$  denotes the position feature vector; $\mathbf{\theta}_{pv}$ denotes the corresponding network training parameters.

\begin{figure}[t]
	\begin{center}
		\includegraphics[width=3.5in]{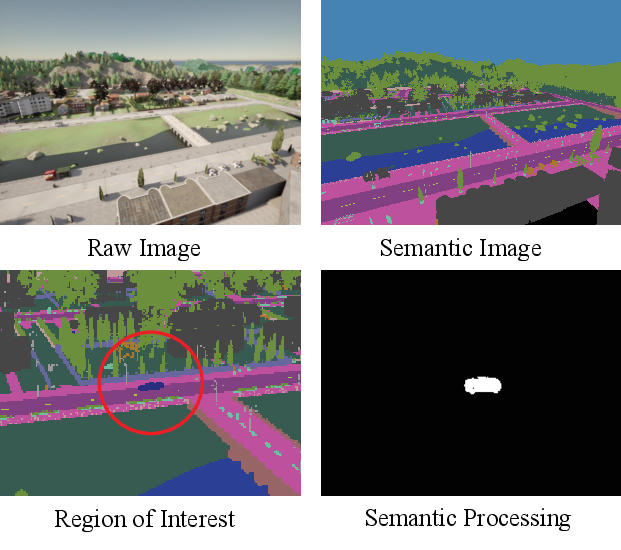}
		\caption{Removing irrelevant features from the ROI.}
		\label{fig1_roi}
	\end{center}
\end{figure}
$\textbf{Image data:}$ As shown in Fig. \ref{fig1_2}(b), the environmental sensing module captures omnidirectional environmental information between the BS and CU. Let the high-definition camera capture images $\mathbf{J} = [\mathbf{j}_{1},\cdots,\mathbf{j}_{N}]\in \mathbb{R}^{N \times h \times w \times 3}$, where $\mathbf{j}_{N}$ denotes the acquisition of the image in the $N$-th communication pair. These images contain spatial information about the CU and key obstacles (e.g., buildings, vegetation, and moving objects) that may affect signal propagation.  To enhance the overall perception of the communication environment, a pretrained residual neural network (ResNet) \cite{targ2016resnet} is employed to extract coarse-grained global scattering features from the raw images.
Specifically, the ResNet backbone $f_{r}(\cdot)$ maps $\mathbf{J}$ into a hierarchical feature representation, in which the original pixel-level information is progressively abstracted through multiple convolutional layers and residual connections, thereby transforming low-level visual features of the environment into high-level semantic-aware representations.
The resulting global environment features are capable of characterizing the overall spatial layout of the communication scenario, the distribution of dominant scatterers, and potential blockage relationships, i.e.,
\begin{equation}
	\begin{aligned}
		\mathbf{J}_{v} =f_{r}(\mathbf{J},\mathbf{\theta}_{jv}),
	\end{aligned}
\end{equation}
where $\mathbf{J}_{v}$  denotes the image feature vector; $\mathbf{\theta}_{jv}$ denotes the corresponding network training parameters.

$\textbf{Semantic data:}$ 
As shown in Fig. \ref{fig1_2}(b), to more accurately characterize the impact of environmental information on SemCom system performance, a refined modeling strategy is adopted for image data. Specifically, semantic segmentation is first performed on raw images $\mathbf{J}$ using PSPNet \cite{zhao2017pyramid} to extract pixel-level semantic features that capture environmental structures and semantics. Based on these representations, a circular ROI mechanism centered around the CU is further introduced according to the CU's two-dimensional image coordinates, through which fine-grained local semantic features closely coupled with signal propagation characteristics are extracted from the segmentation results. Such a mechanism enables the effective capture of local scatterer distributions and near-field blockage effects, thereby facilitating a more accurate characterization of environmental impacts on SemCom system performance.



To account for variations in scatter distribution and spatial complexity across different propagation environments, the ROI radius is designed as an adaptively tunable parameter, denoted by $d_r$, whose value is adjusted according to the density of the surrounding scene. For instance, in urban road scenarios where scatters are sparse, a larger radius (e.g., $d_r$ = 100 pixels) is adopted to cover a broader range of potentially influential regions. In contrast, in forest environments characterized by densely distributed obstacles, a smaller radius (e.g., $d_r$ = 60 pixels) is employed to emphasize the dominant role of near-field surroundings in signal propagation. Furthermore, a predefined semantic filtering threshold is introduced to discard semantic components within the ROI that contribute marginally or are irrelevant to communication performance, thereby mitigating the interference of redundant semantics in the system modeling process.

As illustrated in Fig.~\ref{fig1_roi}, fine-grained local semantic features are extracted from the semantic segmentation results of environmental images. Specifically, PSPNet is adopted to perform pixel-level semantic segmentation, yielding a semantic label map in which each pixel is assigned one semantic category, such as ``road,'' ``vehicle,'' ``building,'' ``vegetation,'' and ``sky.'' Let the pixel-wise semantic label map be denoted by:
\begin{equation}
	L(u,v)\in\{1,2,\ldots,Z\},
\end{equation}
where $L(u,v)$ denotes the semantic label assigned to pixel $(u,v)$, and $Z=28$ represents the total number of semantic classes.
Given the two-dimensional coordinates of the CU on the image plane as $U=(u_0,v_0)$, a circular ROI centered at $U$ with radius $d_r$ is defined as:
\begin{equation}
	\mathcal{F}(U,d_r)=\left\{(u,v)\mid (u-u_0)^2+(v-v_0)^2\le d_r^2 \right\}.
\end{equation}

For each semantic class $z\in\{1,2,\ldots,Z\}$, the corresponding binary mask within the ROI is given by:
\begin{equation}
	B_z(u,v)=\mathbb{I}\!\left[L(u,v)=z\right], \quad (u,v)\in\mathcal{F}(U,d_r),
\end{equation}
where $\mathbb{I}(\cdot)$ denotes the indicator function. Through this binarization process, pixels belonging to semantic category $z$ within the ROI are assigned a value of 1, while all other pixels are assigned with 0. Based on the obtained binary mask, the number of pixels associated with semantic category $z$ inside the ROI is calculated as:
\begin{equation}
	j_{pc,z}=\sum_{(u,v)\in\mathcal{F}(U,d_r)} B_z(u,v).
\end{equation}

Subsequently, $j_{pc,z}$ is compared with a dynamically determined threshold $j_r$ so that semantic categories with marginal contributions to communication performance can be filtered out. The refined semantic feature corresponding to category $z$ is given by:
\begin{equation}
	j_{po,z}=
	\begin{cases}
		0, & j_{pc,z}<j_r,\\
		j_{pc,z}, & j_{pc,z}\ge j_r.
	\end{cases}
\end{equation}

In this way, semantic categories weakly related to propagation characteristics are removed, while local semantic components associated with scattering, blockage, and reflection are preserved. The refined semantic representation is then written as:
\begin{equation}
	\mathbf{j}_{po}=[j_{po,1},j_{po,2},\ldots,j_{po,Z}]^{\mathrm T}.
\end{equation}
\begin{figure}[t]
	\begin{center}
		\includegraphics[width=3.5in]{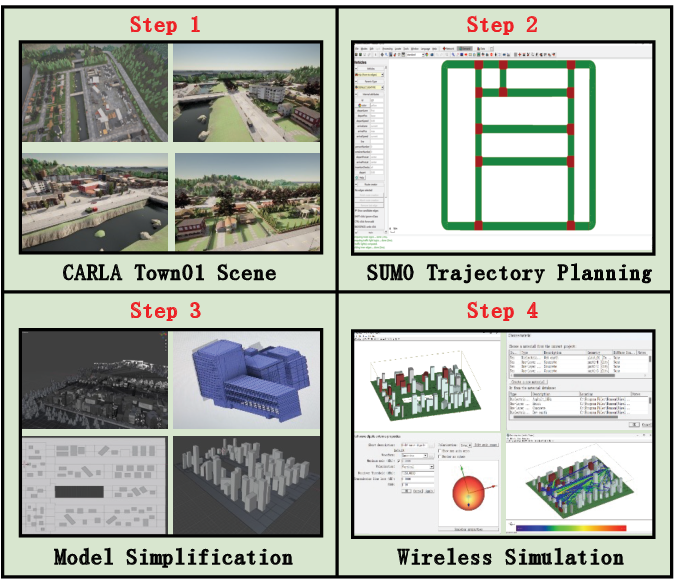}
		\caption{Scene model construction.}
		\label{fig1_4}
	\end{center}
\end{figure}
After ROI-based filtering, the refined semantic feature vector $\mathbf{j}_{po}$ is fed into a vectorization module to obtain a compact latent representation for subsequent learning. Specifically, a stack of FC layers is employed to map $\mathbf{j}_{po}$ into:
\begin{equation}
	\mathbf{J}_{sv}=f_{sv}(\mathbf{j}_{po};\boldsymbol{\theta}_{sv}),
\end{equation}
where $\mathbf{J}_{sv}$ denotes the vectorized semantic feature and $\boldsymbol{\theta}_{sv}$ represents the trainable parameters of the vectorization network.

\textbf{Channel data:} As shown in Fig. \ref{fig1_4}, a 3D propagation environment including the BS and CU is first constructed in Car Learning to Act (CARLA)\cite{dosovitskiy2017carla} and Simulation of Urban Mobility (SUMO)\cite{li2021novel}. The model is then simplified in Blender \cite{soni2023review} by removing scene components that are not relevant to wireless propagation. The processed 3D scene is imported into Wireless InSite (WI)\cite{alkhateeb2019deepmimo} for ray-tracing wireless simulation, from which the channel matrix $\mathbf{H}\in\mathbb{C}^{K\times M}$ under each environmental configuration is obtained. By establishing a one-to-one correspondence between environmental representations and simulated channel responses, a multidimensional environmental dataset is constructed.

\subsection{Channel Knowledge Base Construction}

The multidimensional environmental feature fusion and channel knowledge generation process is described in Fig.~\ref{fig1_2}(c). Specifically, the location feature, global image feature, and fine-grained semantic feature are first projected into a common latent space and then fused through an attention-based multidimensional feature fusion module, which adaptively captures the contributions of heterogeneous environmental factors to channel characteristics. The fused representation is subsequently fed into a Transformer network to learn the mapping from environmental information to channel matrix representation.

\begin{figure}[t]
	\begin{center}
		\includegraphics[width=3.4in]{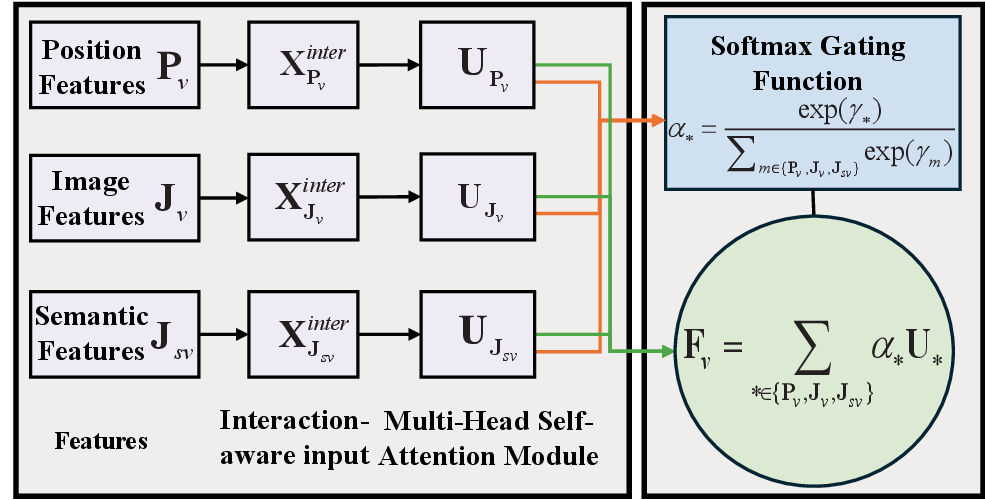}
		\caption{Multidimensional feature fusion.}
		\label{self_t}
	\end{center}
\end{figure}

After multi-source feature extraction, direct summation is generally insufficient to capture the intrinsic correlations and heterogeneous dependencies among different feature dimensions, which limits the expressiveness of the fused representation. To address this issue, an attention-based multidimensional feature fusion mechanism is introduced to enhance feature interactions and adaptively assign importance weights across different environmental dimensions. The specific fusion architecture is illustrated in Fig.~\ref{self_t}.

Let $\mathbf{P}_{v}$, $\mathbf{J}_{v}$, and $\mathbf{J}_{sv}$ denote the extracted location feature, image feature, and semantic feature, respectively. These heterogeneous features are first projected into a unified latent space as:
\begin{equation}
	\mathbf{X}_{m} = f_{m}^{\mathrm{proj}}(\mathbf{E}_{m}), \quad m\in\{p,i,s\},
\end{equation}
where $\mathbf{E}_{p}=\mathbf{P}_{v}$, $\mathbf{E}_{i}=\mathbf{J}_{v}$, and $\mathbf{E}_{s}=\mathbf{J}_{sv}$, while $f_{m}^{\mathrm{proj}}(\cdot)$ denotes the projection function for the $m$-th feature type.

For each projected feature $\mathbf{X}_{m}$, trainable linear transformations are applied to obtain the query, key, and value matrices, i.e.,
\begin{equation}
	\mathbf{Q}_{m}=\mathbf{W}_{m}^{Q}\mathbf{X}_{m}, \quad
	\mathbf{K}_{m}=\mathbf{W}_{m}^{K}\mathbf{X}_{m}, \quad
	\mathbf{V}_{m}=\mathbf{W}_{m}^{V}\mathbf{X}_{m},
\end{equation}
where $\mathbf{W}_{m}^{Q}$, $\mathbf{W}_{m}^{K}$, and $\mathbf{W}_{m}^{V}$ are trainable projection matrices. Then, the attention-enhanced representation of the $m$-th feature is computed as:
\begin{equation}
	\mathbf{U}_{m}
	=
	\mathrm{SoftMax}\!\left(
	\frac{\mathbf{Q}_{m}\mathbf{K}_{m}^{\mathsf T}}{\sqrt{d_k}}
	\right)\mathbf{V}_{m},
\end{equation}
where $d_k$ is the dimension of the key vector.

To further improve the representation capability, a multi-head extension is adopted. For the $r$-th head, the attended output is given by:
\begin{equation}
	\mathbf{U}_{m}^{(r)}
	=
	\mathrm{SoftMax}\!\left(
	\frac{\mathbf{Q}_{m}^{(r)}(\mathbf{K}_{m}^{(r)})^{\mathsf T}}{\sqrt{d_k}}
	\right)\mathbf{V}_{m}^{(r)},
\end{equation}
and the final output is obtained by concatenating all heads followed by a linear projection, i.e.,
\begin{equation}
	\mathbf{U}_{m}
	=
	\mathrm{Concat}\!\left(
	\mathbf{U}_{m}^{(1)},\ldots,\mathbf{U}_{m}^{(R)}
	\right)\mathbf{W}_{m}^{O},
\end{equation}
where $R$ denotes the number of attention heads and $\mathbf{W}_{m}^{O}$ is the output projection matrix.

After obtaining the attention-enhanced representations from different feature dimensions, adaptive weighted fusion is performed as:
\begin{equation}
	\mathbf{F}_{v}
	=
	\sum_{m\in\{p,i,s\}}\alpha_{m}\mathbf{U}_{m},
\end{equation}
where $\alpha_{m}$ denotes the contribution coefficient of the $m$-th feature dimension. Specifically, $\alpha_{m}$ is determined by a softmax-based gating function:
\begin{equation}
	\alpha_{m}
	=
	\frac{\exp(\gamma_{m})}
	{\sum_{n\in\{p,i,s\}}\exp(\gamma_{n})},
	\quad
	\gamma_{m}
	=
	\mathbf{w}^{\mathsf T}\sigma(\mathbf{U}_{m}),
\end{equation}
where $\mathbf{w}$ is a learnable parameter vector and $\sigma(\cdot)$ denotes a nonlinear activation function.

After obtaining the fused environmental representation $\mathbf{F}_{v}$, it is fed into a Transformer network to generate the corresponding channel knowledge representation. The overall mapping is formulated as:
\begin{equation}
	\mathbf{H}_{ne}
	=
	f_{\mathrm{Trans}}\!\left(\mathbf{F}_{v};\boldsymbol{\theta}_{\mathrm{Trans}}\right),
\end{equation}
where $f_{\mathrm{Trans}}(\cdot)$ denotes the Transformer-based mapping function, $\boldsymbol{\theta}_{\mathrm{Trans}}$ represents the trainable parameters of the Transformer, and $\mathbf{H}_{ne}$ denotes the generated environment-aware channel knowledge representation.

During training, the ground-truth channel matrix $\mathbf{H}$ obtained from ray-tracing simulation is used as the supervision target to guide the learning of the environment-to-channel mapping. Correspondingly, the training objective can be written as:
\begin{equation}
	\mathcal{L}_{\mathrm{CKB}}
	=
	\left\|
	\mathbf{H}_{ne}-\mathbf{H}
	\right\|_{2}^{2}.
\end{equation}

Finally, the set of environment-conditioned channel knowledge representations generated from different samples forms the proposed CKB, which provides structured channel priors for subsequent SemCom and JSCC design.

\begin{figure*}[t]
	\begin{center}
		\includegraphics[width=7in]{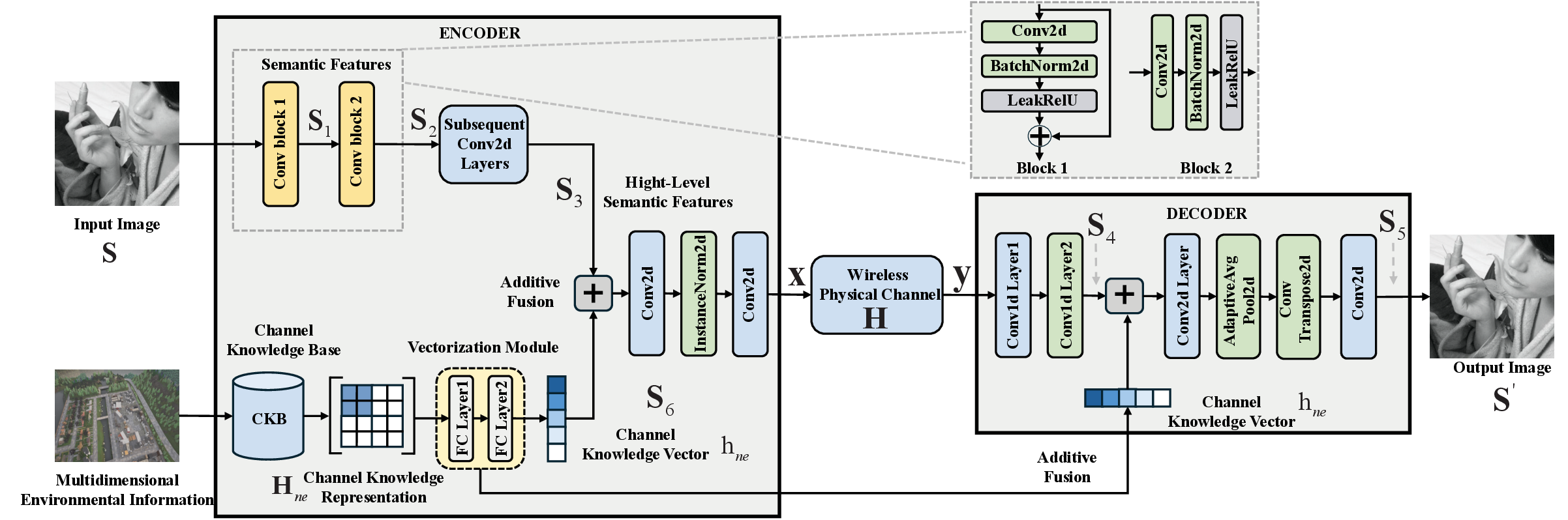}
		\caption{CKB-driven JSCC SemCom system framework.}
		\label{ECKB_DSCAF}
	\end{center}
\end{figure*}
\section{CKB-Driven JSCC SemCom Framework}
The detailed generative CKB-driven JSCC SemCom framework is illustrated in Fig.~\ref{ECKB_DSCAF}. At the transmitter, the SI is processed by a JSCC encoder to extract high-level semantic features, which are fused with the channel knowledge vector retrieved from the CKB to generate an environment-adaptive encoded signal. The signal is then transmitted over the physical channel. At the receiver, the decoder jointly utilizes the received signal and the corresponding channel knowledge to reconstruct the original image.


\subsection{Encoder and CKB-Driven Feature Injection}
At the encoder, $\mathbf{S}$ is initially processed by two convolutional blocks (Conv blocks) with a residual structure to alleviate feature degradation in deep networks and accelerate convergence. Each Conv block consists of two cascaded Conv2d layers. The corresponding feature extraction can be expressed as:
\begin{equation}
	\mathbf{S}_{1}= \delta\!\left(\mathrm{BN}\!\left(\mathcal{E}(\mathbf{S})\right)\right),
\end{equation}
\begin{equation}
	\mathbf{S}_{2}= \delta\!\left(\mathrm{BN}\!\left(\mathcal{E}(\mathbf{S}_{1})\right)\right),
\end{equation}
where $\mathcal{E}(\cdot)$ denotes the Conv block operation, $\mathrm{BN}(\cdot)$ denotes BatchNorm2d, and $\delta(\cdot)$ denotes a nonlinear activation function (e.g., PReLU). The output $\mathbf{S}_{2}$ is further processed by subsequent Conv2d layers to yield a higher-level semantic feature $\mathbf{S}_{3}$.  


Based on the established mapping between environmental information and channel characteristics, the corresponding channel knowledge $\mathbf{H}_{ne}$ is retrieved from the CKB. The retrieved knowledge encapsulates rich multidimensional environmental spatial features and is incorporated into the SemCom system to drive the end-to-end training of the JSCC framework. To align it with the semantic feature space, $\mathbf{H}_{ne}$ is transformed into a vector representation $\mathbf{h}_{ne}$ via a vectorization module composed of two FC layers, where the dimensionality of $\mathbf{h}_{ne}$ is matched to that of $\mathbf{S}_{3}$.


The CKB knowledge is injected into the semantic feature stream through additive fusion followed by convolutional refinement, yielding the final encoded feature vector $\mathbf{x}$:
\begin{equation}
	\mathbf{x}= \mathcal{E}_{2}\Big(\sigma\big(\mathrm{IN}(\mathcal{E}_{1}(\mathbf{S}_{3}+\mathbf{h}_{ne}))\big)\Big),
\end{equation}
where $\mathcal{E}_{1}(\cdot)$ and $\mathcal{E}_{2}(\cdot)$ denote Conv2d operations, $\mathrm{IN}(\cdot)$ denotes InstanceNorm2d, and $\sigma(\cdot)$ denotes an activation function (e.g., ReLU).

\subsection{Physical Channel Transmission and Decoder}
The encoded feature vector $\mathbf{x}$ is transmitted over the physical channel $\mathbf{H}$, producing the received signal $\mathbf{y}$. At the receiver, $\mathbf{y}$ is first processed by two consecutive Conv1d layers to obtain an intermediate feature representation, denoted by $\mathbf{S}_{4}$. Subsequently, $\mathbf{S}_{4}$ is fused with the CKB-derived vector $\mathbf{h}_{ne}$ and passed through a Conv2d layer to generate $\mathbf{S}_{5}$. Finally, $\mathbf{S}_{5}$ is sequentially processed by AdaptiveAvgPool2d, ConvTranspose2d, and Conv2d layers to reconstruct the output image $\mathbf{S}^{'}$.

For clarity, the overall end-to-end mapping of the proposed CKB-driven JSCC SemCom framework can be compactly written as:
\begin{equation}
	\mathbf{S}^{'} = f_{\mathrm{SemCom}}\!\left(\mathbf{S},\mathbf{H}_{ne},\mathbf{H};\boldsymbol{\theta}_{scom}\right),
\end{equation}
where $\boldsymbol{\theta}_{scom}$ denotes the set of trainable parameters.

\subsection{Training Objective}
The proposed framework is trained in an end-to-end manner by minimizing the pixel-wise reconstruction error between the reconstructed image $\mathbf{S}^{'}$ and the original input $\mathbf{S}$. Given a training set $\{(\mathbf{S}^{(i)},\mathbf{H}^{(i)},\mathbf{H}^{(i)}_{ne})\}_{i=1}^{N_{um}}$, the commonly used pixel-level mean-squared-error (MSE) loss is formulated as:
\begin{equation}
	\mathcal{L}_{\mathrm{rec}}(\boldsymbol{\theta})
	=\frac{1}{N_{u}}\sum_{i=1}^{N_{u}}\left\|f_{\mathrm{SemCom}}\!\left(\mathbf{S}^{(i)},\mathbf{H}^{(i)}_{ne},\mathbf{H}^{(i)};\boldsymbol{\theta}\right)-\mathbf{S}^{(i)}\right\|_{2}^{2}.
\end{equation}
where $N_{u}$ denotes the number of training samples, $\mathbf{S}^{(i)}$ represents the SI of the i-th sample, $\mathbf{H}^{(i)}_{ne}$ and $\mathbf{H}^{(i)}$ respectively represent the channel knowledge and channel matrix corresponding to the i-th sample, and $\boldsymbol{\theta}$ denotes the trainable parameter vector.

\section{Simulations}
Our simulation framework leverages multiple software tools, including  CARLA, SUMO, Blender, and WI, to establish a comprehensive 6G-V2X simulation environment. First, we construct the world model using CARLA to create a realistic virtual environment. Subsequently, we design and implement vehicle movement trajectories in SUMO to simulate realistic mobility patterns. The environmental model is then simplified in Blender, where we remove unnecessary building structures while preserving key architectural features. The simplified scene model is then imported into Wireless InSite for electromagnetic simulation. After configuring the relevant electromagnetic parameters, we perform detailed simulations to obtain complete channel characteristics data.

\subsection{Simulation Setup}

\textbf{City Environment Model:} We utilize the autonomous driving simulator CARLA to simulate a street traffic environment, including street scenery, vehicles, and cameras. The proposed scheme is deployed in Town 01, with the BS positioned at point $\mathbf{p}_{u} (x=180, y=240, z=40)$. Three driving routes are planned in the traffic simulation software SUMO. The dimensions of the vehicle are 3.17~m $\times$ 2~m $\times$ 1.5~m (length $\times$ width $\times$ height), and the vehicle speed is set to $3$ m/s. During the simulation, the environment sensing module on the BS adjusts its pitch and azimuth angles in real time to track the vehicle's movement, capturing image data continuously.


\textbf{Electromagnetic Environment Model:} We employ the 3D ray-tracing simulator Wireless InSite to generate the channel matrices between the BS and the vehicle. First, we import the simplified geometric model of Town 01 into Wireless InSite and assign material properties e.g., concrete, brick, wood, and glass to the buildings. In this simulator, the buildings and the size, position, and orientation of the wireless vehicles are kept identical to those in CARLA. Then, we configure the wireless simulation parameters as listed in Table \ref{tabel}, where both the BS and the vehicle are equipped with a $4 \times 4$ antenna array arranged in a uniform planar array distribution.

During the operation of vehicles in CARLA, we record 995 sets of coordinate data (BS and vehicle positions), 995 sets of BS-to-vehicle perspective image data, and 995 sets of semantic data. In Wireless InSite, we set up 995 signal reception points to collect 995 sets of channel matrix data, thereby constructing a wireless environment dataset based on the Town 01 scenario.

\begin{table}[h]
	\centering
	\caption{Wireless Insite Parameters}
	\label{tabel}
	\begin{tabular}{ c|c }
		\hline
		Parameter & Value \\
		\hline
		Communication System & MIMO \\
		Carrier Frequency & 28 GHz \\
		BS Number of antennas &  4$\times$4 \\
		User Number of antennas & 4$\times$4 \\
		Maximum number of signal reflections & 6 \\
		Maximum acceptance path & 20 \\
		\hline
	\end{tabular}
\end{table}

\textbf{Datasets for Validation:} For the evaluation of transfer performance, we employ the CIFAR-10 dataset, which consists of 50,000 images with dimensions of $3\times32\times32$ (color channels, height, width) as the training dataset and 1,000 images as the test dataset  \cite{recht2018cifar}. To further demonstrate the image restoration performance on high-resolution data, we validate the system using the ImageNet \cite{deng2009imagenet} dataset by extracting $32 \times 32$ patches from the center of the images for performance assessment.
\subsection{Evaluation Indicators}
\textbf{MSE:} The  MSE quantifies the error between the generated channel matrix and the ground-truth channel matrix, which is defined as:
\begin{equation}
	\mathrm{MSE}(\mathbf{H}_{ne},\mathbf{H})
	=
	\frac{1}{MK}
	\sum_{i=1}^{M}
	\sum_{j=1}^{K}
	\left(H_{ne}(i,j)-H(i,j)\right)^2 .
\end{equation}

\textbf{PSNR:} The PSNR is derived from the MSE and is widely used to evaluate reconstruction quality in image processing tasks:
\begin{equation}
	\mathrm{PSNR}
	=
	10\log_{10}
	\left(
	\frac{MAX^2}{\mathrm{MSE}(\mathbf{S},\mathbf{S}^{'})}
	\right),
\end{equation}
where $\mathrm{MSE}(\mathbf{S},\mathbf{S}^{'})$ denotes the MSE between the SI $\mathbf{S}$ and the reconstructed image $\mathbf{S}^{'}$, and $MAX$ represents the maximum possible pixel value of the image. A larger PSNR generally indicates higher reconstruction fidelity. However, PSNR does not always correlate well with perceived visual quality.

\textbf{SSIM:} The SSIM is designed to measure perceptual similarity between two images by incorporating structural information of the visual scene. Specifically, SSIM evaluates three key components of images: luminance, contrast, and structural similarity. For two images $\mathbf{S}$ and $\mathbf{S}^{'}$, these components are defined as:
\begin{equation}
	l(\mathbf{S},\mathbf{S}^{'})
	=
	\frac{2\mu_{\mathbf{S}}\mu_{\mathbf{S}^{'}}+c_{1}}
	{\mu_{\mathbf{S}}^{2}+\mu_{\mathbf{S}^{'}}^{2}+c_{1}},
\end{equation}

\begin{equation}
	c(\mathbf{S},\mathbf{S}^{'})
	=
	\frac{2\sigma_{\mathbf{S}}\sigma_{\mathbf{S}^{'}}+c_{2}}
	{\sigma_{\mathbf{S}}^{2}+\sigma_{\mathbf{S}^{'}}^{2}+c_{2}},
\end{equation}

\begin{equation}
	s(\mathbf{S},\mathbf{S}^{'})
	=
	\frac{\sigma_{\mathbf{S}\mathbf{S}^{'}}+c_{3}}
	{\sigma_{\mathbf{S}}\sigma_{\mathbf{S}^{'}}+c_{3}},
\end{equation}
and the overall SSIM is expressed as:
\begin{equation}
	\mathrm{SSIM}(\mathbf{S},\mathbf{S}^{'})
	=
	l(\mathbf{S},\mathbf{S}^{'})^{\alpha}
	c(\mathbf{S},\mathbf{S}^{'})^{\beta}
	s(\mathbf{S},\mathbf{S}^{'})^{\gamma}.
\end{equation}where, $\mu_{\mathbf{S}}$ and $\mu_{\mathbf{S}^{'}}$ denote the mean intensities of the images, $\sigma_{\mathbf{S}}$ and $\sigma_{\mathbf{S}^{'}}$ denote the corresponding standard deviations, and $\sigma_{\mathbf{S}\mathbf{S}^{'}}$ represents the covariance between $\mathbf{S}$ and $\mathbf{S}^{'}$. The constants $c_i$, $i\in\{1,2,3\}$, and $\alpha$, $\beta$, $\gamma$ are positive parameters used to stabilize the division operations.

\begin{figure}[t]
	\begin{center}
		\includegraphics[width=3.5in]{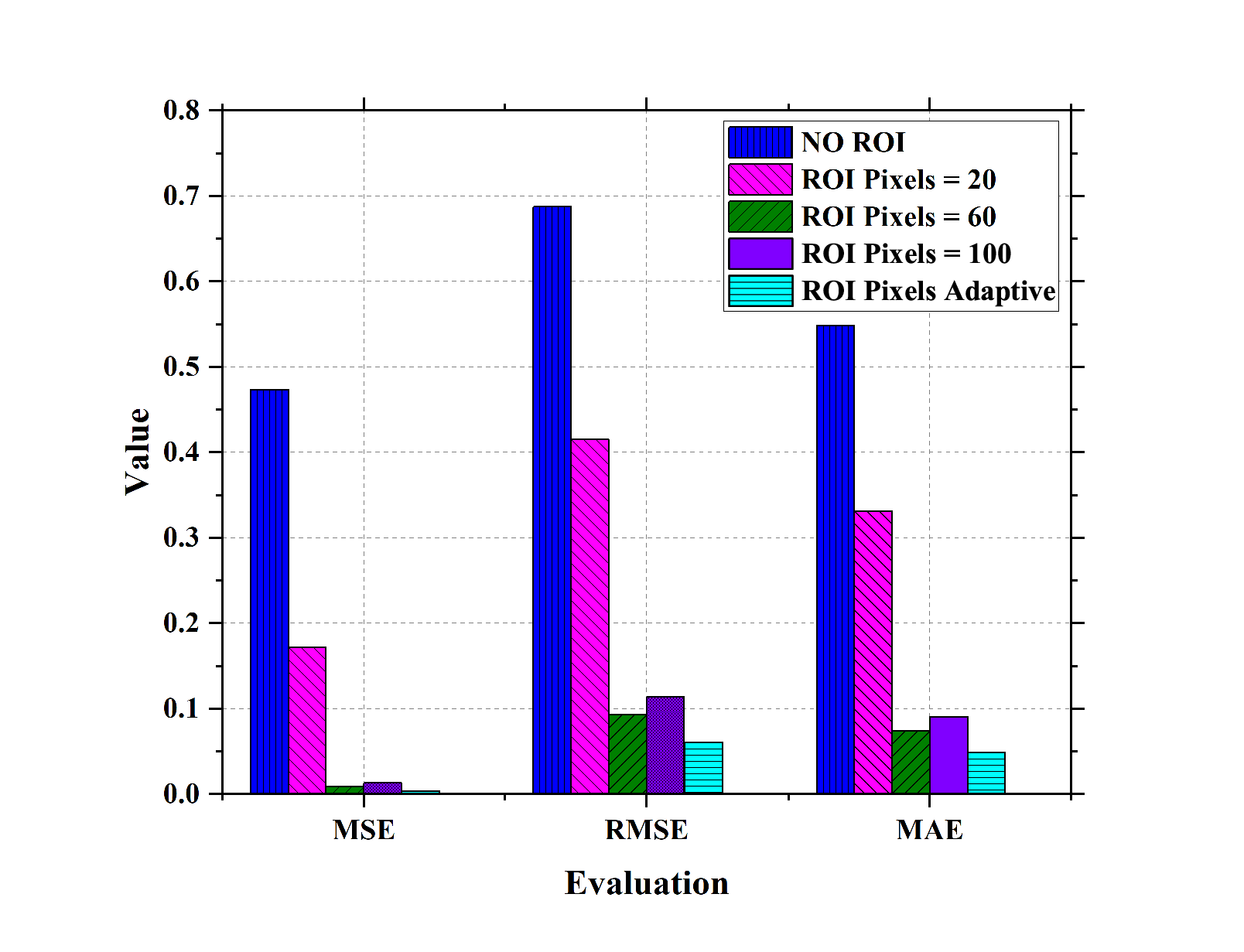}
		\caption{Channel generation accuracy under different ROI pixels.}
		\label{fig_exp_5}
	\end{center}
\end{figure}
\begin{figure}[t]
	\begin{center}
		\includegraphics[width=3.5in]{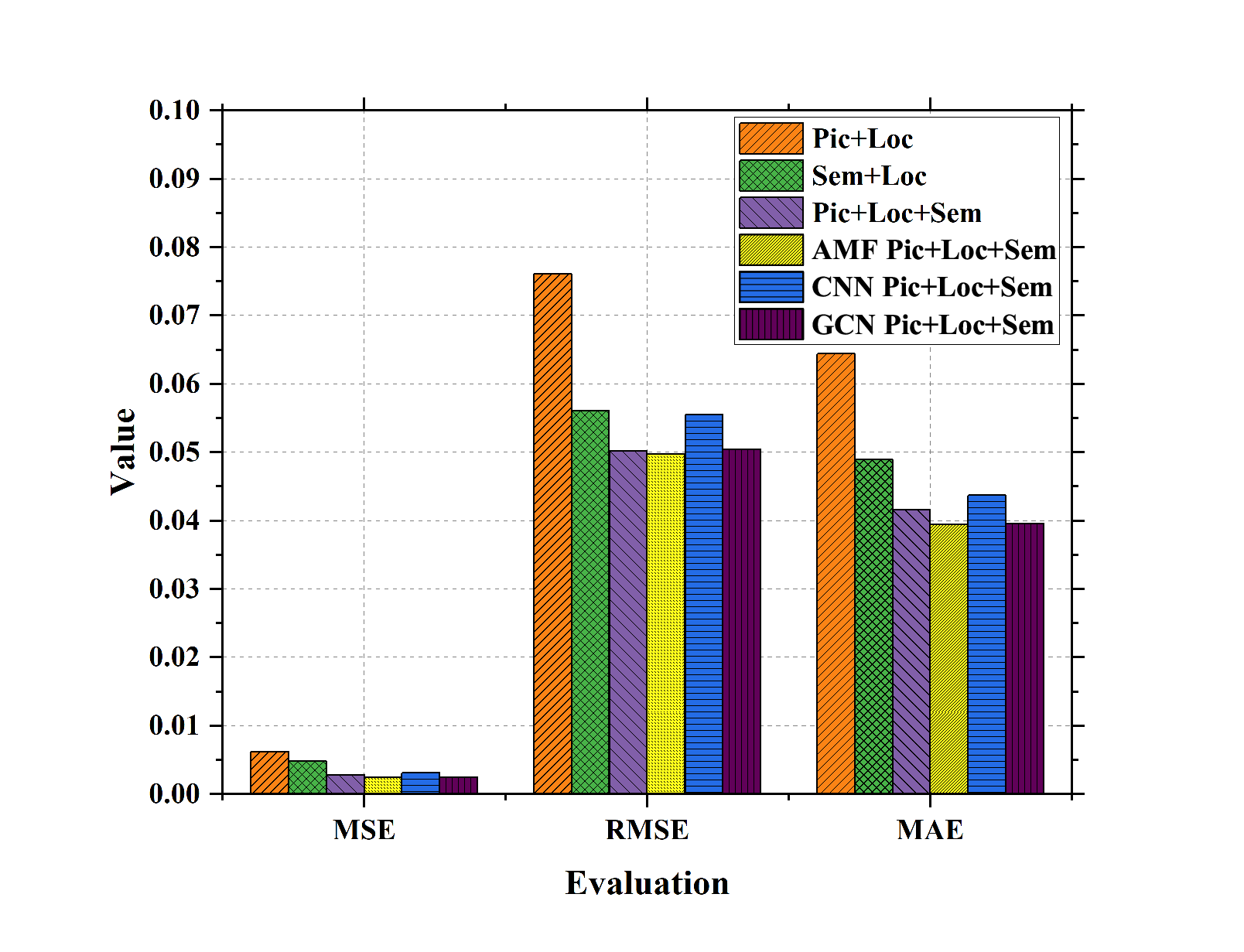}
		\caption{Channel generation accuracy under different feature fusion methods.}
		\label{fig_exp_1}
	\end{center}
\end{figure}
\begin{figure}[t]
	\begin{center}
		\includegraphics[width=3.5in]{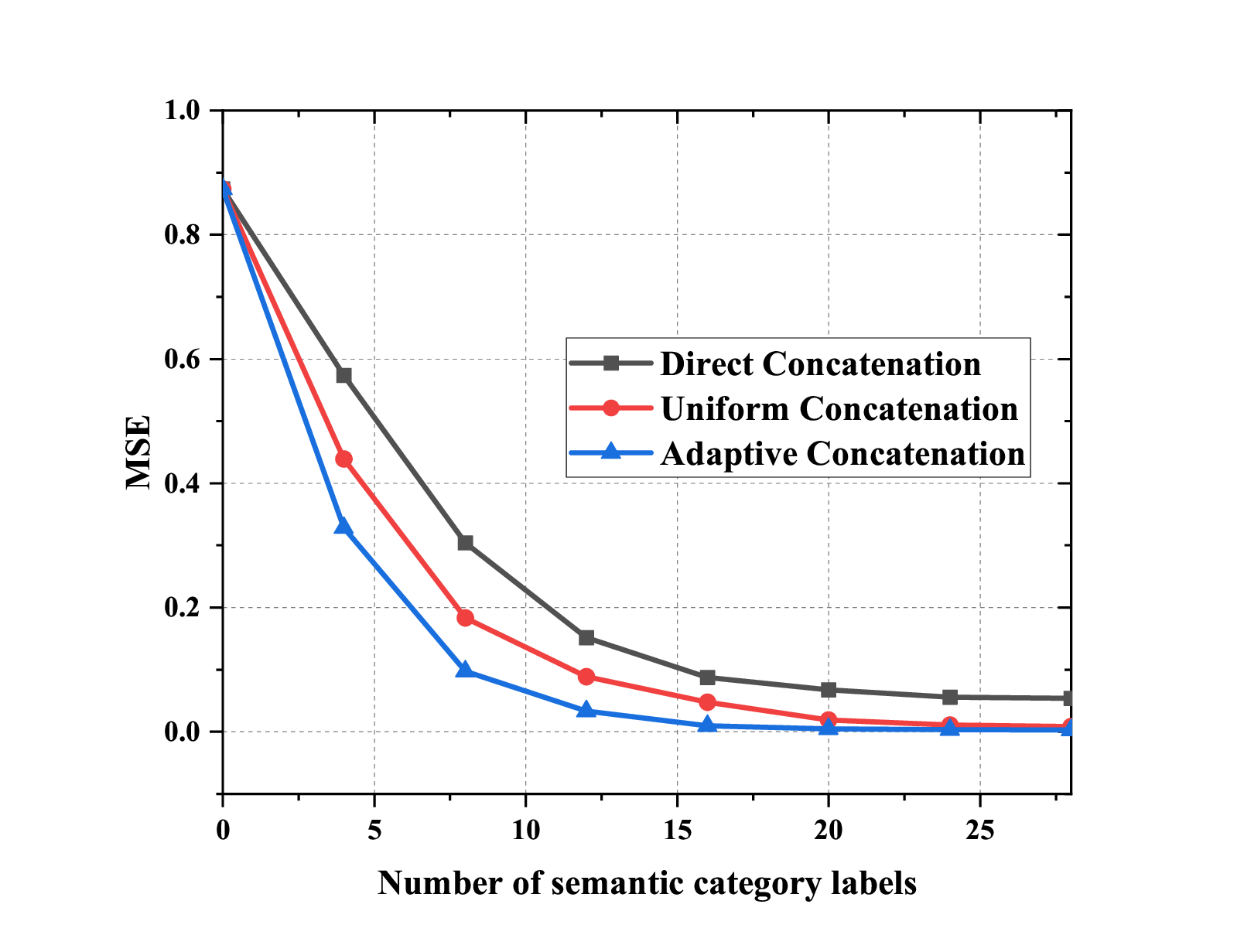}
		\caption{Channel generation accuracy under different numbers of semantic category labels.}
		\label{fig_exp_6}
	\end{center}
\end{figure}
\begin{figure}[t]
	\begin{center}
		\includegraphics[width=3.5in]{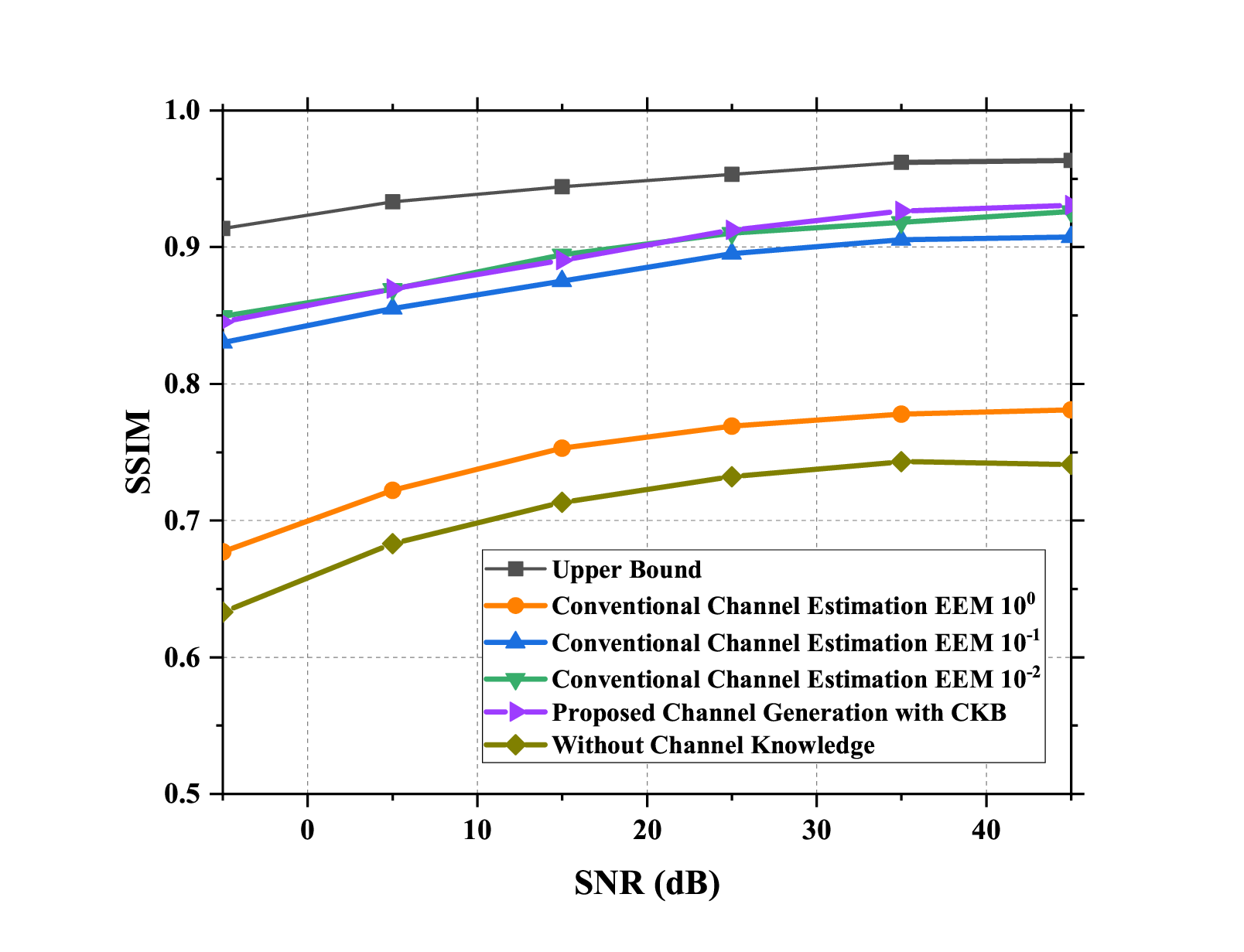}
		\caption{SSIM under different channel generation schemes.}
		\label{fig_exp_2}
	\end{center}
\end{figure}
\subsection{ROI Pixels}
We evaluate the channel knowledge generation performance with different ROI pixel configurations. Specifically, 995 groups of multidimensional environmental features and their corresponding channel matrix data are randomly partitioned into the training and test sets with a ratio of 3:1. The results are presented in Fig.~\ref{fig_exp_5}. When no ROI is applied and only global image features are utilized, the channel matrices generated from the multidimensional environmental information exhibit the poorest accuracy. When the ROI pixel size is set to 20, 60, and 100, the resulting MSE values are 0.1721, 0.0087, and 0.0129, respectively, indicating that increasing the ROI size does not monotonically improve the generation performance. Furthermore, when the ROI pixel size is adaptively adjusted in response to scene variations, all three evaluation metrics, i.e., MSE, root mean square error (RMSE), and mean absolute error (MAE), are consistently improved compared with those obtained using a fixed ROI pixel size. These results demonstrate that, in the dynamic 6G-V2X scenario, an adaptive ROI configuration is more effective in capturing the mapping relationship between multidimensional environmental information and channel matrices, thereby enabling the generation of more accurate channel knowledge.

\subsection{Channel Generation}
We evaluate the performance of channel generation with multidimensional environmental information fusion. As illustrated in Fig.~\ref{fig_exp_1}, the MSE of the proposed SemCom environment-based channel generation scheme reaches $10^{-3}$. Comparative experiments further reveal that channel generation methods relying solely on 2D environmental semantics perform significantly worse in terms of MSE, RMSE, and MAE when compared to methods incorporating 3D feature fusion. To further explore the efficacy of feature fusion strategies, we evaluate four representative approaches: linear fusion, self-attention-based adaptive multimodal fusion (AMF), nonlinear fusion based on convolutional neural networks (CNNs), and graph convolutional network (GCN)-based fusion optimization. Among these, the self-attention-based method achieves the lowest MSE of 0.0026, outperforming all other fusion schemes. Thus, the self-attention-based fusion mechanism further enhances inter-feature dependencies and adaptively adjusts fusion weights, leading to superior channel representation quality and robustness.

\subsection{Number of Semantic Category Labels}
We evaluate the impact of the number of semantic category labels and the contribution coefficients $\alpha_{*}$ of each attended feature in the attention-based feature fusion mechanism on the channel knowledge generation performance. In the experiments, multidimensional features are fused using three strategies, including direct concatenation, uniform concatenation, and adaptive concatenation of $\alpha_{*}$. The experimental results are illustrated in Fig.~\ref{fig_exp_6}. As the number of semantic category labels increases from 0 to 28, the MSE of the generated channel matrices decreases rapidly at first and then gradually converges, which indicates that a sufficient number of semantic labels is required during semantic feature extraction to characterize the real environment. However, increasing the number of semantic categories does not yield a significant improvement in channel matrix generation accuracy. In particular, the optimal MSE performance is achieved when 28 semantic categories are utilized. Moreover, with different numbers of semantic label settings, the channel matrices generated with adaptive $\alpha_{*}$ consistently yield the lowest MSE. These results demonstrate that adopting 28 semantic categories together with an adaptive $\alpha_{*}$ configuration enables a more accurate mapping between multidimensional environmental information and channel matrices.
\begin{figure}[t]
	\centering
	\subfloat[ \rmfamily SSIM indicator]{
		\includegraphics[scale=0.35]{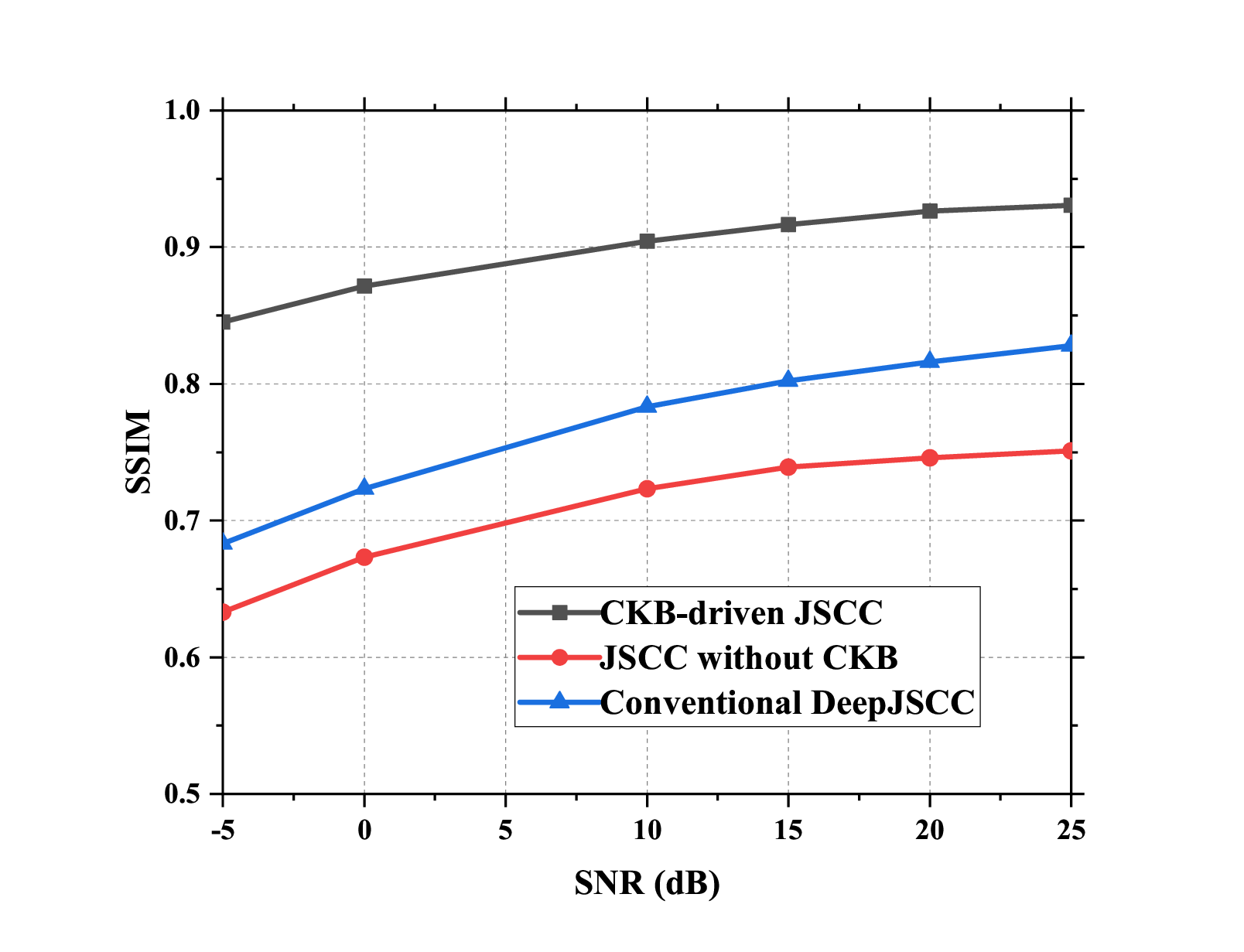}
		\label{fig:exp_4_1}
	}
	\qquad
	\subfloat[\rmfamily PSNR indicator]{
		\includegraphics[scale=0.35]{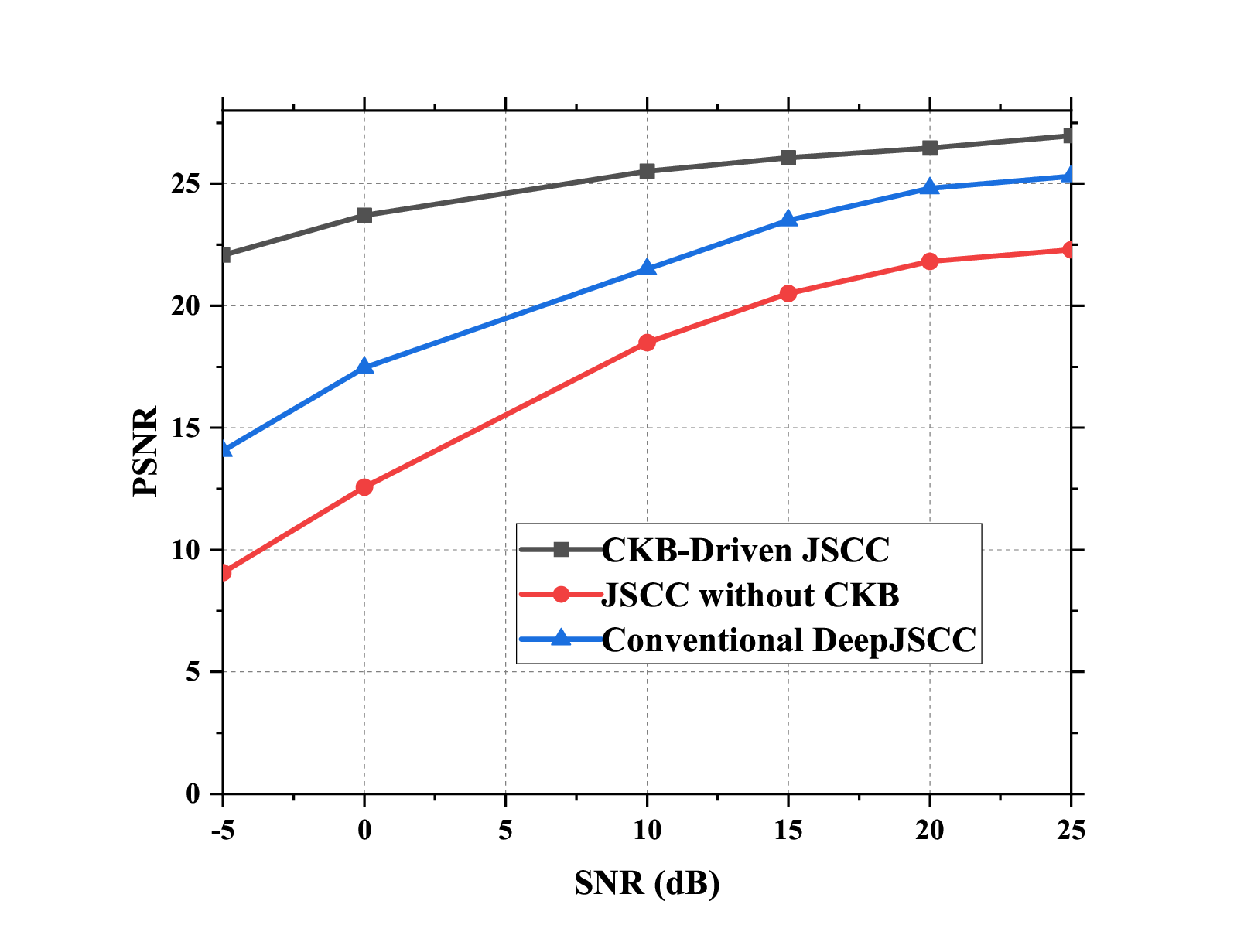}
		\label{fig:exp_4_2}
	}
	\caption{Performance comparison of SemCom systems. (a) SSIM indicator. (b) PSNR indicator.}
	\label{fig:exp_4_x}
\end{figure}

\subsection{Structural Similarity Index}
We evaluate the performance of the CKB in a SemCom system using the SSIM metric to assess image recovery. As shown in Fig.~\ref{fig_exp_2}, when the channel knowledge input to the joint encoder-decoder module matches the physical channel matrix, the SSIM value is the highest, resulting in optimal image recovery, which is considered as the theoretical upper bound. The channel knowledge generated based on CKB exhibits an estimation error with $10^{-3}$. Even with low SNR, satisfactory SSIM values are obtained. We then compare the channel knowledge in CKB with traditional channel estimation. Note that the estimation error magnitude of conventional channel estimation methods exhibits dynamic variation. When the channel estimation error magnitude (EEM) reaches $10^{-2}$, the image recovery performance improves with increasing SNR, and the SSIM value approaches that of the SemCom system with generated channel knowledge. However, as the error magnitude increases, the image recovery performance gradually decreases. It is also observed that the image recovery performance of the SemCom system with channel knowledge in the joint encoder-decoder module is superior to the system without channel knowledge.

\begin{figure*}[t]
	\begin{center}
		\includegraphics[width=7in]{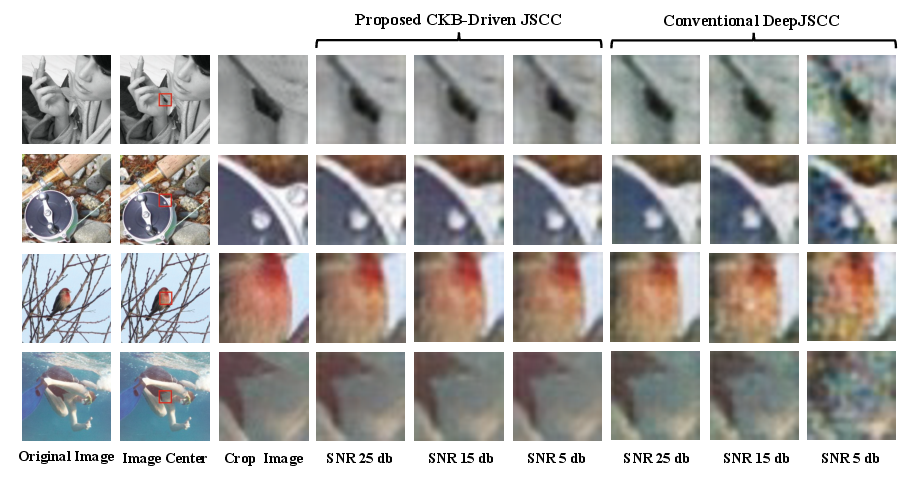}
		\caption{Image reconstruction.}
		\label{fig_exp_3}
	\end{center}
\end{figure*}
\subsection{CKB-Driven JSCC SemCom System}
We compare the transmission performance of the proposed CKB-driven JSCC SemCom system with the SemCom system without CKB-driven JSCC and the conventional DeepJSCC system \cite{bourtsoulatze2019deep}. As shown in Fig.~\ref{fig:exp_4_x}, in real physical channels impaired by environmental factors, the semantic system without CKB demonstrates inferior SSIM and PSNR performance (although DeepJSCC slightly outperforms this baseline). However, in the CKB-driven JSCC SemCom system, both the PSNR and SSIM performance metrics show significant improvements, maintaining $\mathrm{PSNR} > 22.073$ dB and $\mathrm{SSIM} > 0.84$ across the $[-5, 25]$ dB SNR range, indicating that this CKB-driven JSCC SemCom system framework demonstrates promising transmission reliability.


\subsection{Image Reconstruction}
We demonstrate the image reconstruction performance using high-resolution images and validate the results on the ImageNet dataset. From this dataset, we extract $32 \times 32$ pixel images at the center of the original images for transmission in the SemCom system, where the generated channel knowledge is incorporated into the joint encoder--decoder architecture of the SemCom system. As shown in Fig.~\ref{fig_exp_3}, the conventional DeepJSCC method achieves reasonable image restoration performance at high SNR regimes. However, the restoration performance gradually degrades as the SNR decreases. In contrast, the CKB-driven JSCC SemCom system maintains excellent image restoration capability with both high and low SNR, and the image reconstruction quality only declines a little as the SNR decreases. Additionally, we do not retrain the model on the ImageNet dataset, but we still attain good image recovery results.


\section{Conclusion}
In this paper, a generative CKB is developed for 6G-V2X SemCom systems to capture the mapping between multidimensional environmental information and channel characteristics. Based on the constructed CKB, a CKB-driven JSCC SemCom framework is proposed to incorporate environment-aware channel knowledge into the semantic communication process.﻿
Simulation results demonstrate that the proposed method achieves channel-generation error with $10^{-3}$ MSE and significantly improves transmission performance compared with conventional DeepJSCC schemes and JSCC without CKB. In particular, the incorporation of environment-aware channel knowledge effectively mitigates channel estimation errors and enhances system robustness in complex propagation environments.﻿
Furthermore, the results reveal that multidimensional environmental feature modeling and adaptive feature fusion play a critical role in accurately characterizing channel dynamics and improving SemCom performance.
Future work will focus on extending the proposed framework to more dynamic scenarios and reducing the dependency on large-scale environment-aware datasets.
﻿
\bibliography{CKB}
\end{document}